\documentclass[useAMS,usenatbib,usegraphicx]{mn2e}
\usepackage{amsmath,multirow,color}

\title[Obscured star formation in LABs]{Obscured star formation in Ly$\bmath{\alpha}$ blobs at $\bmath{z = 3.1}$}
\author[Y. Tamura et al.]{Y. Tamura,$^{1}$%
\thanks{E-mail: ytamura@ioa.s.u-tokyo.ac.jp}
Y. Matsuda,$^{2,3}$ %
S. Ikarashi,$^{1}$ %
K.~S. Scott,$^{4}$ %
B. Hatsukade,$^{5}$ %
\newauthor
H. Umehata,$^{1}$ %
T. Saito,$^{6}$ %
K. Nakanishi,$^{3,7,8}$ %
M.~S. Yun,$^{9}$ %
H. Ezawa,$^{3}$ %
\newauthor
D.~H. Hughes,$^{10}$ %
D. Iono,$^{11,3}$ %
R. Kawabe,$^{3,7}$ %
K. Kohno,$^{1}$ %
G.~W. Wilson$^{9}$
\\
$^{1}$Institute of Astronomy, The University of Tokyo, Mitaka, Tokyo 181-0015, Japan\\
$^{2}$California Institute of Technology, MC 105-24, 1200 East California Boulevard, Pasadena, CA 91125, USA\\
$^{3}$National Astronomical Observatory of Japan, Mitaka, Tokyo 181-8588, Japan\\
$^{4}$North American ALMA Science Center, National Radio Astronomy Observatory, Charlottesville, VA 22903, USA\\
$^{5}$Department of Astronomy, Kyoto University, Kyoto 606-8502, Japan\\
$^{6}$Kavli Institute for the Physics and Mathematics of the Universe, The University of Tokyo, Kashiwanoha, Kashiwa, 277-8583, Japan\\
$^{7}$Joint ALMA Observatory, Alonso de Cordova 3107, Vitacura, Santiago 763 0355, Chile\\
$^{8}$The Graduate University for Advanced Studies (Sokendai), Mitaka, Tokyo 181-8588, Japan\\
$^{9}$Department of Astronomy, University of Massachusetts, Amherst, MA 01003, USA\\
$^{10}$Instituto Nacional de Astrof\'{i}sica, \'{O}ptica y Electr\'{o}nica (INAOE), Aptdo.\ Postal 51 y 216, 72000 Puebla, Pue., Mexico\\
$^{11}$Nobeyama Radio Observatory, National Astronomical Observatory of Japan, Minaminaki, Minamisaku, Nagano 384-1305, Japan
}
\begin{document}

\date{Accepted 2013 January 11. Received 2013 January 9; in original form 2012 May 30}

\pagerange{\pageref{firstpage}--\pageref{lastpage}} \pubyear{2012}

\maketitle

\label{firstpage}

\begin{abstract}
We present results from the AzTEC/ASTE 1.1-mm imaging survey of 35 Ly$\alpha$ blobs (LABs) found in the SSA22 protocluster at $z = 3.1$. These 1.1-mm data reach an r.m.s.\ noise level of 0.7--1~mJy~beam$^{-1}$, making this the largest millimetre-wave survey of LABs to date. 
No significant ($\ge 3.5\sigma$) emission is found in any of individual 35 LABs, and from this, we estimate 3$\sigma$ upper limits on the far-infrared luminosity of $L_{\rm FIR} < 2 \times 10^{12} L_{\odot}$. 
Stacking analysis reveals that the 1.1-mm flux density averaged over the LABs is $S_{\rm 1.1mm} < 0.40$~mJy (3$\sigma$), which places a constraint of $L_{\rm FIR} < 4.5 \times 10^{11} L_{\odot}$. 
This indicates that earlier 850-$\micron$ measurements of the LABs may have overestimated their flux densities.
Our results suggest that LABs on average have little ultra-luminous obscured star-formation, in contrast to a long-believed picture that LABs undergo an intense episode of dusty star-formation activities with star-formation rates of $\sim 10^3 M_{\sun}$~yr$^{-1}$. Observations with ALMA are needed to directly study the obscured part of star-formation activity in the LABs. 
\end{abstract}



\section{Introduction}

Ly$\alpha$ blobs (LABs) are characterized by extended (20--300~kpc) Ly$\alpha$ nebulae that are often found in overdense regions at high redshift. The origin of Ly$\alpha$ nebulosity, however, is mysterious. 
There are possible explanations for the origin: 
The scenario that was first proposed is that the Ly$\alpha$ nebulae are produced by mechanical feedback (or `superwind') or photo-ionisation from active galactic nuclei (AGN) and/or massive star-formation activities \citep{Taniguchi00, Taniguchi01, Ohyama03, Mori06}.  In fact, ultraviolet (UV) continuum and/or 24-\micron\ emission, the latter arising from starburst/AGN heating of dust, are often detected in LABs \citep{Steidel00, Matsuda04}, which can provide the sufficient number of ionising photons \citep{Webb09, Geach09, Colbert11} to account for the Ly$\alpha$ luminosities \citep[$L_{\rm Ly\alpha} \ga 10^{42.5}$~erg~s$^{-1}$, e.g.,][]{Matsuda04, Matsuda11, Saito06, Saito08}.  The large velocity width of the Ly$\alpha$ emission \citep[$\sim$550 km~s$^{-1}$,][]{Matsuda06} can also be accounted for by the superwind scenario. 
On the other hand, a sizable number of LABs which lack evidence of such apparent heating sources have been reported.  This fact imposes  an alternative scenario in which the origin of Ly$\alpha$ nebulae is attributed to cooling radiation from primeval hydrogen gas which accretes on to massive dark haloes \citep[a.k.a.\ cold accretion; e.g.,][]{Fardal01, Nilsson06, Smith08}, although there remains the possibility that the ionising sources are hidden by the interstellar medium (ISM) located along the line of sight. 

\begin{figure*}
\includegraphics[scale=0.66,angle=-90]{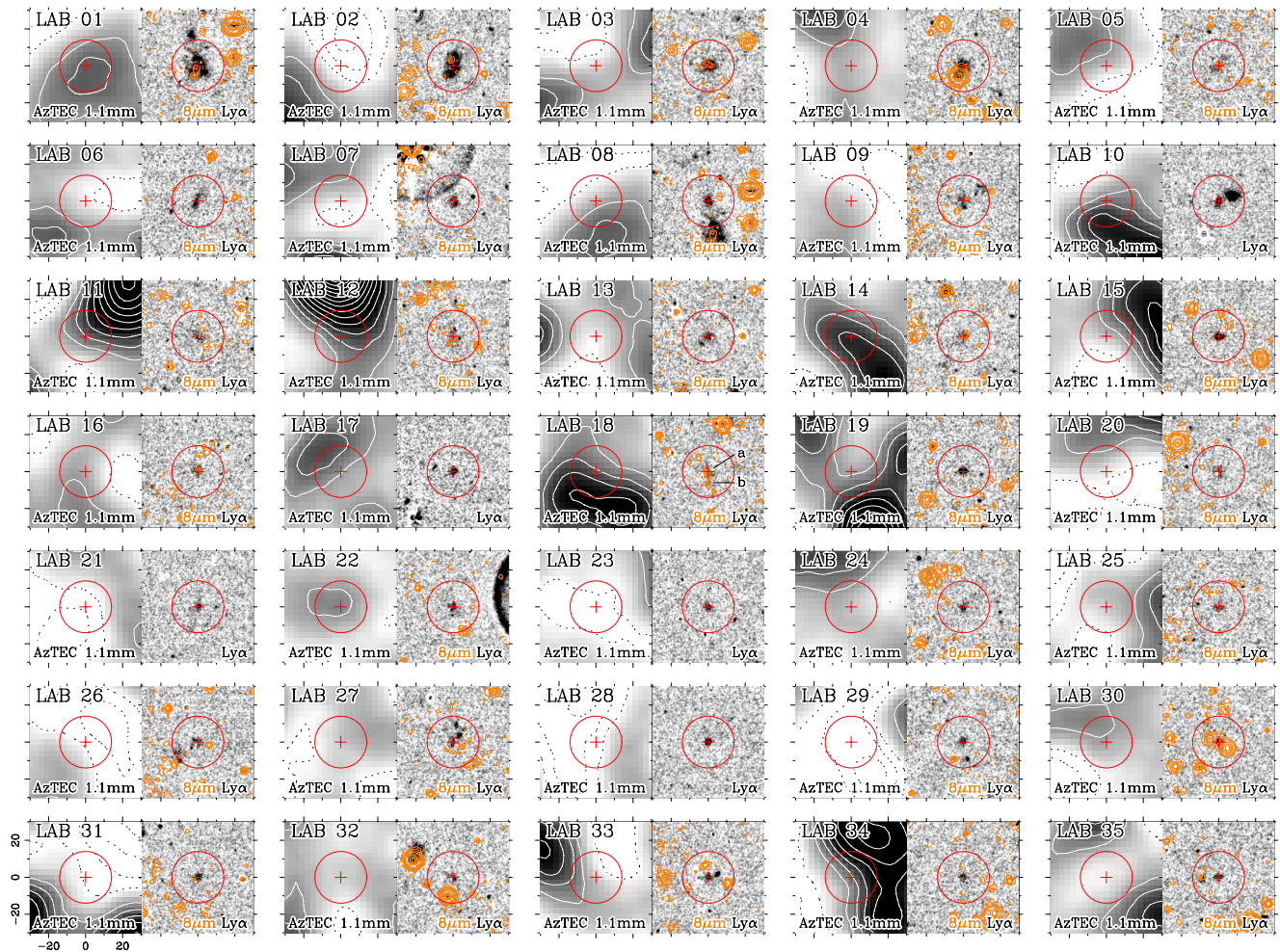}
 \caption{The left-hand panel: The 1.1-mm postage stamp images ($60'' \times 60''$) for all 35 LABs. The contours start from 1$\sigma$ with an interval of 1$\sigma$. The negative signals are indicated by dotted contours. The $1\sigma$ noise levels are 0.7--1 mJy~beam$^{-1}$, depending on locations across the 1.1 mm image. 
 The right-hand panel: The Subaru ${\rm NB497} - BV$ images, which indicate the intensity of Ly$\alpha$ emission at $z = 3.1$.  The orange contours show IRAC 8~$\micron$ \citep{Webb09}, which are drawn at (2, 5, 10, 20, 50, 100, ...) times local noise levels ($1\sigma \approx 1 \times 10^{-2}$ MJy~str$^{-1}$ in typical).}
 \label{fig:postagestamp}
\end{figure*}

Observations of obscured star-formation and/or AGN are therefore necessary to properly understand the origins of the Ly$\alpha$ nebulosity.  Many attempts to detect the interstellar cold dust and molecular gas in LABs at millimetre (mm) and sub-mm wavelengths have been carried out \citep{Chapman01, Chapman04, Geach05, Matsuda07, Beelen08, Yang12}.  However, whether LABs have intense star-formation activities that are capable of producing and maintaining the Ly$\alpha$ haloes is still controversial.

In this paper, we present the results from our unbiased 1.1-mm survey of 35 LABs at $z = 3.1$ found in optical narrow-band filter observations \citep{Steidel00, Matsuda04} toward the SSA22 field, which is known for having an overdensity of Ly$\alpha$ emitters (LAEs) at $z= 3.09$ \citep{Hayashino04}.  
This is the largest mm survey of LABs to date, for which we can study the obscured star formation of these systems.
%
The structure of this paper is as follows. In \S~2, we describe our 1.1-mm observations and data reduction. \S~3 describes the results.  Finally, we have brief discussions and a summary in \S~4. 
Throughout this paper, we assume a concordance cosmology with $\Omega _{\rm m} = 0.3$, $\Omega _{\Lambda} = 0.7$, $H_0 = 70$~km~s$^{-1}$~Mpc$^{-1}$, where 1$''$ corresponds to a physical scale of 7.64~kpc at $z = 3.09$. 


\section{Observations}

The data were taken with the AzTEC 1.1-mm camera \citep{Wilson08} installed on ASTE \citep{Ezawa04} located at Pampa la Bola, Atacama desert, Chile. The data taken during July--September 2007 is described in \citet{Tamura09}. In addition to the 2007 data, we added new data taken in 2008 that almost tripple the survey area to 0.27~deg$^2$. The complete description will be given elsewhere (Tamura et al., in preparation).

The reduction procedure is described in \citet{Scott08} and \citet{Downes12}. The time-stream data were intensively cleaned using a principal component analysis (PCA) algorithm, and then mapped. The full width at half maximum (FWHM) of the point response function is 34$''$, corresponding to 260~kpc in physical scale at $z = 3.1$. The pointing was checked every 1~hr.  Uranus and Neptune were used for flux calibration, yielding an absolute accuracy better than 10 percent.  The resulting r.m.s.\ noise over the region covering 0.27~deg$^2$ is 0.7--1.2~mJy beam$^{-1}$ ($\le 0.8$~mJy beam$^{-1}$ for 30 out of the 35 LABs). Note that stacking analysis for \textit{Spitzer}/MIPS, IRAC, and VLA sources in SSA22 shows no systematic error in astrometry down to better than 4$''$.  Submillimeter Array (SMA) 860-$\micron$ imaging of the brightest 1.1-mm source, SSA22-AzTEC1 \citep{Tamura10}, also supports this.

\begin{table}
 \centering
 \begin{minipage}{70mm}
  \caption{The 1.1-mm properties of LABs in SSA22.}
  \begin{tabular}{@{}lcccc@{}}
  \hline
   Name  & \multicolumn{3}{c}{1.1 mm results} & Other results\\
                & $S_\rmn{1.1mm}$ & $\sigma$ & S/N 
				& $S_\rmn{850\mu m}$\footnote{Observed by SCUBA \citep{Geach05}. The LABs detected at 850 \micron\ with $\ge 3.5\sigma$ are indicated in bold-face type.} \\
                & (mJy) & (mJy) &       & (mJy) \\
\hline
 LAB1\footnote{The $3\sigma$ upper limits of $S_{\rm 880\mu m} < 4.2$~mJy \citep{Matsuda07}, $S_{\rm 870\mu m} < 12$~mJy and $S_{\rm 1.2mm} < 0.45$~mJy \citep{Yang12} are reported.} & 1.97 & 0.74 & 2.7  & $\mathbf{16.8 \pm 2.9}$  \\
 LAB2 & $-1.89$ & 0.76 & $-2.4$ & $3.3 \pm 2.9$  \\
 LAB3 & $-0.69$ & 0.73 & $-0.9$ & $-0.2 \pm 1.2$  \\
 LAB4 & 0.11 & 0.74 & 0.1 & $0.9 \pm 1.5$  \\
 LAB5 & 0.34 & 0.74 & 0.5 & $\mathbf{5.2 \pm 1.5}$  \\
 LAB6 & 0.07 & 1.14 & 0.1 & $-0.5 \pm 1.4$  \\
 LAB7 & $-0.88$ & 0.74 & $-1.2$ & $0.2 \pm 1.6$  \\
 LAB8 & 0.67 & 0.74 & 0.9 & $0.3 \pm 5.3$  \\
 LAB9 & 0.07 & 0.74 & 0.1 & $1.3 \pm 5.3$  \\
 LAB10& 1.20 & 0.84 & 1.4 & $\mathbf{6.1 \pm 1.4}$ \\
 LAB11& 0.61 & 0.73 & 0.8 & $-0.4 \pm 5.3$ \\
 LAB12& 0.30 & 0.74 & 0.4 & $3.2 \pm 1.6$ \\
 LAB13& $-0.72$ & 0.73 & $-1.0$ & ... \\
 LAB14& 2.43 & 0.76 & 3.2 & $\mathbf{4.9 \pm 1.3}$ \\
 LAB15& $-0.27$ & 0.74 & $-0.4$ & ... \\
 LAB16& 0.34 & 0.74 & 0.5 & $2.2 \pm 5.3$ \\
 LAB17& 1.41 & 1.19 & 1.2 & ... \\
 LAB18-a& 1.53 & 0.73 & 2.1 & \multirow{2}{*}{$\Bigr\}\,\mathbf{11.0 \pm 1.5}$} \\
 LAB18-b& 2.33 & 0.73 & 3.2 & \\
 LAB19& $-0.81$ & 0.74 & $-1.1$ & $-8.6 \pm 5.3$ \\
 LAB20& $-0.80$ & 0.75 & $-1.1$ & $0.4 \pm 1.5$ \\
 LAB21& $-1.37$ & 0.75 & $-1.8$ & ... \\
 LAB22& 1.04 & 0.74 & 1.4 & ... \\
 LAB23& $-1.55$ & 0.80 & $-1.9$ & ... \\
 LAB24& 0.03 & 0.72 & 0.0 & ... \\
 LAB25& 0.01 & 0.73 & 1.4 & $1.4 \pm 5.3$ \\
 LAB26& $-0.90$ & 0.74 & $-1.2$ & $-2.7 \pm 5.3$ \\
 LAB27& 0.18 & 0.77 & 0.2 & $0.5 \pm 1.6$ \\
 LAB28& $-0.99$ & 0.76 & $-1.3$ & ... \\
 LAB29& $-2.54$ & 0.91 & $-2.8$ & ... \\
 LAB30& $0.65$ & 0.74 & 0.9 & $3.3 \pm 1.3$ \\
 LAB31& $-1.44$ & 0.74 & $-1.9$ & $-3.7 \pm 5.3$ \\
 LAB32& $-0.16$ & 0.74 & $-0.2$ & $1.8 \pm 1.4$ \\
 LAB33& 0.04 & 0.73 & 0.1 & $1.6 \pm 1.5$ \\
 LAB34& 1.01 & 0.93 & 1.1 & ... \\
 LAB35& $-0.74$ & 0.73 & $-1.0$ & $1.2 \pm 5.3$ \\
 \hline
Mean & $< 0.40$\footnote{The $3\sigma$ upper limit.} & ... & ... & $3.0 \pm 0.9$\\
\hline
\end{tabular}
\end{minipage}
\end{table}
\begin{figure}
	\includegraphics[scale=0.31,angle=-90]{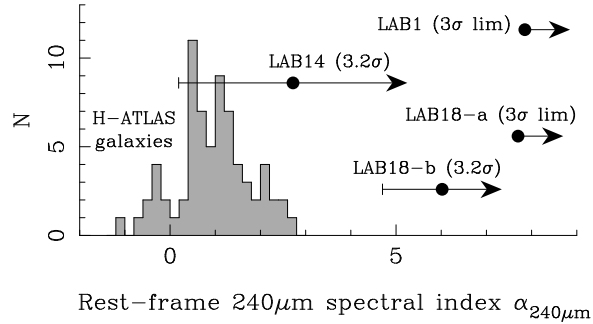}
	\caption{
		Constraint on the rest-frame 240-$\micron$ spectral indices of LABs at $z= 3.1$.
		The histogram shows $\alpha_{\rm 240\mu m}$ found in seventy low-$z$ \textit{Herschel}-ATLAS SDP sources with spectroscopic redshift of $z = 0.20$--0.30, at which the \textit{Herschel}/SPIRE 250 and 350~$\micron$ bands observe the $\approx 240~\micron$ part of SEDs in the rest frame.
		The spectral index of LAB14 is consistent with those of H-ATLAS galaxies.
		However, indices of LAB1 and LAB18-a/b cannot be explained.
	}\label{fig2_alpha}
\end{figure}


\section{Results}
In this section, we first discuss tentative detections of 1.1~mm emission from individual LABs in \S~3.1.  We then consider a statistical detection of the average 1.1~mm properties of the LABs in \S~3.2

\subsection{1.1 mm emission of individual LABs}
We do not find significant ($\ge 3.5\sigma$) 1.1-mm emission for any of the 35 LABs, as shown in Figure~\ref{fig:postagestamp} and Table~1, which lists the 1.1-mm flux density measured at the locations of the LABs.  
Although the peak of Ly$\alpha$ emission may not always coincide with the 1.1-mm counterpart, the offset can be negligible because the Ly$\alpha$ extent is well within the beam ($34''$).
If we assume a dust temperature of $T_{\rm dust} = 35$~K and a dust emissivity index of $\beta = 1.5$, the 3$\sigma$ upper limit places a constraint on far-infrared (FIR) luminosity of $L_{\rm FIR} < 2 \times 10^{12} L_{\odot}$ for the LABs.  This limit corresponds to a star-formation rate (SFR) of $\approx 400 M_{\sun}$~yr$^{-1}$, which suggests that LABs do not have intense dust-obscured star-formation activity found in sub-mm galaxies \citep[SMGs,][for a review]{Blain02}.  Given that our 1.1-mm map reveals $> 100$ SMGs over the SSA22 region (Tamura et al., in prep.), none of which coincide with the LABs, this result strongly suggests that the LAB population is essentially different from the SMG population.

We note that SPIRE/\textit{Herschel} data that have recently been taken toward SSA22 (P.I.: Y.\ Matsuda) are in good agreement with the 1.1-mm results.  The 35 LABs have no SPIRE 500-$\micron$ counterpart.  While low-S/N 250 $\micron$ enhancements are seen at the positions of a few LABs, the flux densities rapidly dim toward longer wavelengths, implying that the dust emission seen at 250~$\micron$ is due to high dust temperatures and/or low-$z$ contaminants.  However, identification of exact 250 $\micron$ counterparts is beyond the scope of the present paper.

  In the rest of this section, we discuss three tentative ($>2\sigma$) detections of the 1.1~mm emission from three of the LABs.

\textit{SSA22-LAB1} --- LAB1 was originally discovered by an optical narrow-band filter survey towards SSA22 \citep{Steidel00}, and is one of the most-studied LABs in the mm and sub-mm. Subsequent imaging and photometric observations with the  Submillimetre Common-User Bolometer Array \citep[SCUBA,][]{Holland99} on the James Clerk Maxwell Telescope (JCMT) had revealed a luminous 850-\micron\ source at the position of LAB1 with $S_{\rm 850\mu m} = 16.8 \pm 2.9$~mJy \citep{Chapman01, Chapman04, Geach05}.  However, the SMA 880-\micron\ imaging found no emission, suggesting that the spatial extent of the sub-mm emission of LAB1 should be larger than $4''$ \citep{Matsuda07}.  Very recently, \citet{Yang12} have reported a non-detection of mm and sub-mm emission, suggesting that there is no dusty starburst associated with the LAB as reported by \citet{Chapman01, Chapman04} and \citet{Geach05}.  Our new 1.1-mm map shows only a marginal enhancement of 1.9~mJy~beam$^{-1}$ (2.7$\sigma$) relative to the noise.  These low resolution, single-dish observations cast doubt on the presence of an extended dust component that could account for the SMA non-detection.  

Furthermore, the 850-to-1100~\micron\ flux ratio would be $> 8$, which is quite high compared with a typical starburst galaxies.  The 850--1100~$\micron$ band corresponds to the rest-frame wavelengths of 210--270~$\micron$ for a $z = 3.1$ object.  So, the 1.1-mm $3\sigma$ upper limit of 2.2~mJy places a constraint on the rest-frame spectral index at $\lambda_{\rm rest} = 240~\micron$\footnote{This defines the slope of a spectrum such that $S_{\nu} \propto \nu^{\alpha_{\rm 240\mu m}}$.} to $\alpha_{\rm 240\mu m} = 7.85$ or higher.
In Figure~\ref{fig2_alpha} we show a histogram of $\alpha_{\rm 240\mu m}$ measured in 70 (ultra-)luminous infrared galaxies (U/LIRGs) with spectroscopic redshifts of $z$ = 0.2--0.3.  The U/LIRGs are catalogued in the \textit{Herschel}-ATLAS Science Demonstration Phase \citep[SDP,][]{Eales10,Pascale11,Rigby11,Smith11} database\footnote{www.h-atlas.org/public-data/.}, and all detected at 250 and 350~$\micron$ at $>5\sigma$.  For $z$ = 0.2--0.3 objects, the SPIRE 250--350~$\micron$ bands sample the rest-frame $\approx 240~\micron$ part of the SEDs.  The mean H-ATLAS spectral index inferred from the 250-to-350~$\micron$ flux ratios is $\alpha_{\rm 240\mu m} = 0.93 \pm 0.82$ (the error bar is from the standard deviation), which turns to be lower than expected in the Rayleigh-Jeans regime (This is simply because we are looking at the waveband close to the dust emission peak).
The spectral index of LAB1 is extremely steep compared with the H-ATLAS indices, suggesting that the earlier SCUBA measurement remarkably overestimates the 850~$\micron$ flux density.  On the other hand, our result is consistent with other recent non-detections with the SMA, LABOCA/APEX, and PdBI \citep{Matsuda07, Yang12}.

\emph{SSA22-LAB14} --- The $3.2\sigma$ enhancement is seen at the location of LAB14 (see Fig.~\ref{fig:postagestamp}), which is $\approx$25$''$ north-eastward from the SMG, SSA22-AzTEC69 (S/N = 4.1, Tamura et al.\ in preparation).  The 1.1-mm flux density at the LAB14 position is $2.43 \pm 0.76$~mJy beam$^{-1}$ although heavy blending with SSA22-AzTEC69 makes it difficult to accurately measure the 1.1-mm flux density.  Note that it is unlikely that SSA22-AzTEC69 is the mm counterpart to LAB14 since a Monte Carlo simulation \citep[the method is given in][]{Scott08} shows a low probability ($p \la 0.01$) that a S/N = 4 source is detected $>20''$ away from its original position.
LAB14 has been detected at 850~$\micron$ \citep[SMM~J221735.84+001558.9, $S_{\rm 850\mu m} = 4.9\pm1.3$~mJy,][]{Chapman05, Geach05}.  The 850-to-1100~\micron\ flux ratio would be 2.0 if assuming $S_{\rm 1.1mm} =  2.43 \pm 0.76$~mJy.  
This yields $\alpha_{\rm 240\mu m} = 2.72 \pm 2.53$, which is consistent with those found in the H-ATLAS galaxies (Fig.~\ref{fig2_alpha}), although the 1.1-mm flux is tentative.

\emph{SSA22-LAB18} --- 
A SCUBA detection has been reported for this LAB \citep[$S_{\rm 850\mu m} = 11.0 \pm 1.5$~mJy,][]{Geach05}.  It has two IRAC counterparts, LAB18-a and LAB18-b \citep{Webb09}.  The former coincides with the Ly$\alpha$ peak and has a 24 $\micron$ counterpart, whereas the latter has a hard X-ray source \citep{Geach09} but no 24-$\micron$ counterpart.
We find an enhancement of 1.5~mJy (2.1$\sigma$) and 2.3~mJy (3.2$\sigma$) at the positions of LAB18-a and b, respectively, but the two objects are likely blended by a nearby 1.1-mm source with S/N $\approx$ 4, located $\approx 20''$ south of LAB18-a (or $\approx 10''$ south of LAB18-b). 
The 850-to-1100~\micron\ flux ratio of LAB18-a is $> 7.2$ if taking the 3$\sigma$ upper limit, while that of LAB18-b is $4.7 \pm 1.6$ if the flux density would be $S_{\rm 1.1mm} = 2.33 \pm 0.73$~mJy, although the source blending likely boosts the 1.1-mm flux density.  The spectral indices at $\lambda_{\rm rest} = 240~\micron$ are $> 7.7$ and $6.0 \pm 1.3$ for LAB18-a and b, respectively.  Again, the $\alpha_{\rm 240\mu m}$ indices are substantially deviated from the H-ATLAS distribution (Fig.~\ref{fig2_alpha}), implying that the SCUBA measurement might overestimate the 850-$\micron$ flux.  
Note that the southernmost 1.1-mm source is not likely to be the counterpart because the Monte Carlo simulation suggests a low probability ($p \la 0.15$).

\begin{figure}
		\includegraphics[scale=0.33,angle=-90]{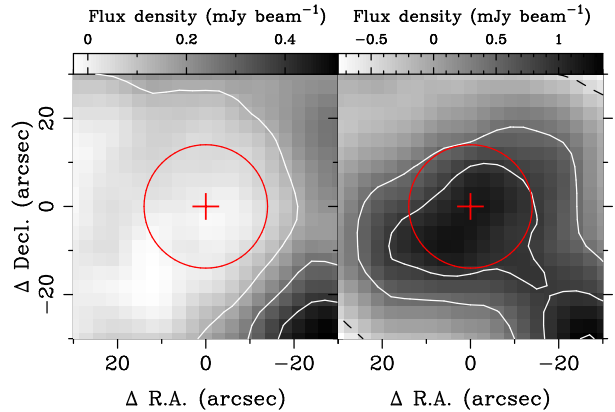}
		 \caption{ The 1.1-mm stacked images at the positions of 32 LABs (Left) and 3 SCUBA-detected LABs (Right). The 1$\sigma$ noise levels at 1.1~mm are 0.135~mJy~beam$^{-1}$ and 0.44~mJy~beam$^{-1}$, respectively. The contours start from 1$\sigma$ with an interval of $1\sigma$, and the negative signals are indicated by dotted contours. The red circle and cross on each panel indicate the HPBW of AzTEC/ASTE and the nominal position of LABs. No significant emission was found in both samples, although a small peak is seen (2.3$\sigma$) in the stacked image of the SCUBA-detected LABs.}
 \label{fig:stack}
\end{figure}

\subsection{Stacking analysis}
Stacking analysis, a pixel-to-pixel weighted-mean of 2-dimensional images around objects of interest, is often used to statistically detect very faint emission features that are common among the objects.  In order to measure the average 1.1-mm flux density of LABs, we stack the 1.1-mm images around the positions of (i) all of the LABs in SSA22, and (ii) the five SCUBA-detected LABs, for which \citet{Geach05} have reported positive detections at 850~$\micron$. 
Note that only LABs that are $> 30''$ away from any of mm-bright ($\ge 3.5\sigma$) point sources (Tamura et al.\ in preparation) are considered to eliminate the blending of the nearby bright sources; this leaves 32 (91 percent) of the 35 LABs\footnote{LAB14, 18 and 34 are masked.} and 3 of the 5 SCUBA-detected LABs\footnote{LAB14 and 18 are masked.}. 
The PCA cleaning process used in AzTEC reduction filters out low spatial frequency components of the map, resulting in axisymmetric negative sidelobes ($\approx -7$ percent of the maximum) around a bright source. The sidelobes systematically offset the zero point of a stacked image. In this analysis, we first deconvolved the 1.1-mm image with a point response function \citep[details are given in][]{Downes12} using the CLEAN algorithm \citep{Hogbom74}. 
The CLEAN-ed images that are cut out around the positions of the 32 LABs are weighted according to the local noise level, and then averaged.  The 1$\sigma$ noise level is estimated by calculating $(\sum_{i} \sigma_{i}^{-2})^{-1/2}$, where $\sigma_{i}$ is the local r.m.s.\ noise level of the 1.1-mm image around the position of the $i$-th LAB.  
We verify that the average (i.e., stacked) flux density of model sources is correctly reproduced by Monte Carlo simulations in which 32 model point sources are placed in the CLEAN-ed image and then the image is stacked at the positions of those model sources (Ikarashi et al., in preparation).

In Figure~\ref{fig:stack} (left panel) we show the results of the stacking analysis for the 32 LABs; the mm emission is not statistically detected.  The weighted mean of the 1.1-mm flux density constrains the typical 1.1-mm flux density, and thus the $L_{\rm FIR}$, for LABs.  We put the 3$\sigma$ upper limit of $S_{\rm 1.1mm} < 0.40$~mJy, which corresponds to $L_{\rm FIR} < 4.5 \times 10^{11} L_{\odot}$ and $M_{\rm dust} < 1 \times 10^8 M_{\odot}$ if assuming $T_{\rm dust} = 35$~K, $\beta = 1.5$ and the dust emissivity $\kappa_{\rm d}(850\micron) = 0.1$~m$^2$~kg$^{-1}$ \citep{Hildebrand83}.
As shown in Figure~\ref{fig2_alpha}, a realistic $\alpha_{\rm 240\mu m}$ is likely in the range between $-1$ and 3, which makes the 850-to-1100~$\micron$ flux ratio of 0.8 to 2.2.  The 1.1-mm $3\sigma$ upper limit thus corresponds to 0.3--0.9~mJy at 850~$\micron$.  This is below the mean 850~$\micron$ flux density of all the LABs observed by SCUBA \citep[$3.0 \pm 0.9$~mJy,][]{Geach05}, but is still consistent with a mean 850~$\micron$ flux of $1.2\pm 0.4$~mJy derived only for the LABs which are not individually detected at 850~$\micron$ \citep{Geach05}.
The right panel of Figure~\ref{fig:stack} shows the 1.1-mm stacked image for the SCUBA-detected LABs.  The noise level is 0.44~mJy~beam$^{-1}$.  We do not significantly detect 1.1-mm emission in the SCUBA-LABs, however, we see a small 2.3$\sigma$ peak.  We derive a 3$\sigma$ upper limit of $S_{\rm 1.1mm} < 3.3$~mJy, yielding $L_{\rm FIR} < 1.4 \times 10^{12} L_{\odot}$ and $M_{\rm dust} < 3 \times 10^8 M_{\odot}$ if assuming $T_{\rm dust} = 35$~K, $\beta = 1.5$ and $\kappa_{\rm d}(850\micron) = 0.1$~m$^2$~kg$^{-1}$. 

\begin{figure}
	\includegraphics[angle=-90,scale=0.33]{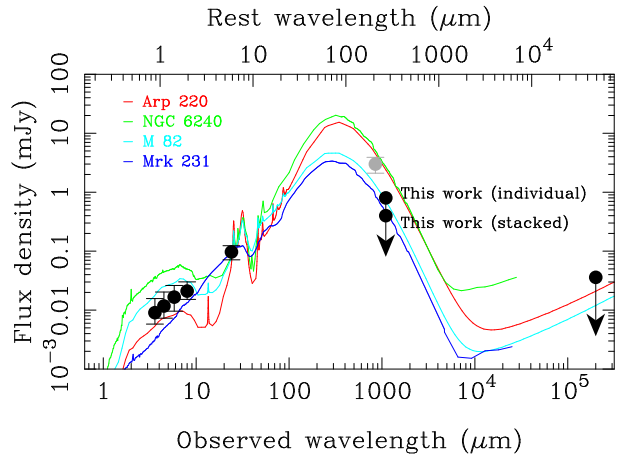}
	\caption{
		The composite SED of the 24-$\micron$--detected LABs \citep[LAB1, LAB14, LAB16, and LAB18,][]{Webb09}. 
		The filled circles and error bars of IRAC and MIPS photometry (3.6--24~$\micron$) represent the mean and minimum-maximum of the flux densities of the four LABs.  We also show the averaged 850-$\micron$ flux (grey circle) and a VLA 21-cm 3$\sigma$ upper limit.
		The template SEDs are normalized by the mean 24-$\micron$ flux of these LABs.
	}\label{fig4_sed}
\end{figure}

\section{Discussions and conclusions}

We have conducted 1.1-mm observations with AzTEC/ASTE to map the SSA22 field, which is known for having an overdensity of $z = 3.1$ LABs, as well as LAEs.  None of the individual 35 LABs have been detected at 1.1 mm, though LAB14 has a marginal signal ($3.2\sigma$).  Our stacking analysis for 32 LABs fails to statistically detect the 1.1~mm emission ($S_{\rm 1.1mm} < 0.40$~mJy, 3$\sigma$), suggesting that LABs on average have little ultra-luminous obscured star-formation ($L_{\rm FIR} < 4.5 \times 10^{11} L_{\odot}$ [3$\sigma$], if assuming $T_{\rm dust} = 35$~K and $\beta = 1.5$), unlike a long-believed picture that many LABs undergo intense dusty star-formation with SFRs of $\sim 10^3 M_{\sun}$~yr$^{-1}$ \citep{Chapman01, Chapman04, Geach05}. 

We compile the results of previous mm/sub-mm observations of LABs ($>30$~kpc) at various redshifts \citep[][and this work]{Smail03, Greve07, Matsuda07, Beelen08, Saito08, Smith08, Ouchi09, Bussmann09, Yang12, Walter12}, and find that the detection rate of mm and sub-mm emission in individual LABs is 4/48 (8.3 percent) \citep[][for sub-mm--detected LABs]{Smail03, Greve07, Beelen08, Yang12} though the sensitivities are not uniform.   This value is lower than previously suggested \citep[5/25 = 20 percent,][]{Geach05}, but at least a small fraction ($\sim$10 percent) of LABs may undergo obscured starbursts.  
Although the bulk of LABs appear not to have starbursts as seen in SMGs, massive (10$^{10}$--10$^{11} M_{\sun}$) stellar components are broadly seen within the Ly$\alpha$ haloes \citep{Geach07, Uchimoto08, Smith08, Ouchi09}.  

Moreover, 4 of 26 (15 percent) and 5 of 29 (17 percent) of the LABs in SSA22 have 24~$\micron$ and X-ray sources, respectively \citep{Webb09, Geach09}, suggesting that 15--20 percent of LABs may host obscured star-formation and/or AGN activities, regardless of whether they are detected at 1.1~mm. 
Figure~\ref{fig4_sed} shows the composite mid-IR to radio SED of the 24-$\micron$ detected LABs \citep[LAB1, LAB14, LAB16, and LAB18-a,][]{Webb09}.  Two of them (LAB14 and LAB18) are detected in the X-rays \citep{Geach09}.  We also show SEDs of local starburst galaxies Arp~220, NGC~6240, M~82 \citep{Silva98}, and a nearby IR-luminous quasar Mrk~231 \citep{Berta05}.  The FIR luminosities of Arp~220, NGC~6240, M~82 and Mrk~231 are $L_{\rm FIR} = 1.4 \times 10^{12} L_{\sun}$, $5.4 \times 10^{11} L_{\sun}$, $4.1 \times 10^{10} L_{\sun}$ and $2.0 \times 10^{12} L_{\sun}$ \citep{Sanders03}, respectively.  The template SEDs are redshifted to $z = 3.09$ and normalized by the mean 24-$\micron$ flux of the four LABs.  M~82 and Mrk~231 have warmer dust than Arp~220 and NGC~6240, and this is why the (sub-)mm fluxes of the M~82 and Mrk~231 templates are lower than the others.  The 1.1-mm upper limits are better consistent with the extrapolation of the M82 and Mrk~231 SEDs than Arp 220 and NGC6240.  This suggests that the 24-$\micron$ objects within the four LABs are powered by star formation and/or AGN activities that are enough to maintain the dust temperatures high, but lack a large reservoir of cooler gas and dust which is often seen in SMGs \citep[$M_{\rm dust} \sim 10^9 M_{\sun}$, e.g.,][]{Kovacs06}.

These evidences may imply that some LABs are at a phase where the extreme starburst phase has just been quenched for some reason, for example, by dissociation of molecular clouds by a superwind from a nuclear starburst and/or AGN.  On the other hand, $\sim 30$ percent of LABs do not host any bright UV continuum sources in the halo \citep[e.g.,][]{Matsuda04, Nilsson06}; such LABs without UV continuum sources may result from cooling radiation of cold streams as suggested by many authors \citep[e.g.,][]{Nilsson06}.

Although the non-detections reported here put a strong constraint on the obscured SFR of the LABs, they do not rule out any possibilities for the formation mechanisms of Ly$\alpha$ nebulosity.  
If all of the Ly$\alpha$ emission observed in the LABs is attributed to ionising photons from young massive stars, the Ly$\alpha$ luminosities correspond to SFRs of  $\approx 10$--100 $M_{\odot}$\,yr$^{-1}$ following the expression $L_{\rm Ly\alpha} = 1.0 \times 10^{42}\,({\rm SFR}/M_{\odot}\,{\rm yr}^{-1})$ erg~s$^{-1}$ \citep{Osterbrock89, Kennicutt98}.  Our constraint on the FIR luminosity ($L_{\rm FIR} < 4.5 \times 10^{11} L_{\odot}$) suggests that SFR obscured by dust is less than 80~$M_{\odot}$\,yr$^{-1}$, following \citet{Kennicutt98}.  This limit is comparable to the Ly$\alpha$-derived SFR, but is not small enough to fully rule out the possibility that the Ly$\alpha$ nebulosity is produced by feedback from massive star-formation activity.  
\citet{Smith08} claimed that their non-detection of 1.2~mm emission in a $z=2.8$ LAB ($L_{\rm Ly\alpha} = 2.1 \times 10^{43}$~erg s$^{-1}$), which limits the SFR to $< 220$~$M_{\sun}$~yr$^{-1}$ (assuming $T_{\rm dust} = 35$~K and $\beta = 1.5$), rules out the photoionisation scenario in favor of the cold accretion scenario.  We consider, however, that the interpretation still leaves room for reconsideration, since only an SFR of 21~$M_{\sun}$~yr$^{-1}$ is able to produce the Ly$\alpha$ luminosity of the $z=2.8$ LAB and so the SFR limit ($< 220 M_{\sun}$~yr$^{-1}$) from the 1.2-mm measurement is not enough to exclude the photoionisation scenario.


Obviously, one of the reasons why the formation mechanism of LABs is so ambiguous is that we do not have a complete picture of obscured star-formation activity within LABs.  The sensitivity of the AzTEC/ASTE imaging survey presented in this work is confusion limited, and higher resolution imaging with higher sensitivity such as possible with ALMA is needed to give a better understanding of the formation mechanism of LABs.  


\section*{Acknowledgments}
We would like to acknowledge to the AzTEC/ASTE team who made the observations possible. 
We thank T.\ Yamada and T.\ Hayashino for providing the Subaru images.  We would also like to thank R.\ Ivison for providing the new VLA image.
YT is supported by JSPS Grant-in-Aid for Research Activity Start-up (no.\ 23840007).  
KSS is supported by the National Radio Astronomy Observatory, which is a facility of the National Science Foundation operated under cooperative agreement by Associated Universities, Inc.
BH is supported by Research Fellowship for Young Scientists from JSPS. 
AzTEC/ASTE observations were partly supported by KAKENHI (no.\ 19403005, 20001003).
The ASTE project is driven by Nobeyama Radio Observatory (NRO), a branch of NAOJ, in collaboration with University of Chile, and Japanese institutes including University of Tokyo, Nagoya University, Osaka Prefecture University, Ibaraki University and Hokkaido University.
The \textit{Herschel}-ATLAS is a project with \textit{Herschel}, which is an ESA space observatory with science instruments provided by European-led principal investigator consortia and with important participation from NASA. The H-ATLAS website is http://www.h-atlas.org/.

\bsp

\label{lastpage}

\end{document}